\documentclass{optica-article}

\journal{opticajournal} 

\articletype{Research Article}

\usepackage{lineno}

\usepackage{xr}
\usepackage{soul}
\usepackage{amsmath}
\usepackage{graphicx}
\usepackage{epstopdf}
\usepackage{subfig}
\usepackage{multirow}
\usepackage{url}
\usepackage{float}
\usepackage{bm}
\usepackage{ifthen}
\usepackage{braket}

\usepackage{siunitx} 

\usepackage[absolute]{textpos}
\usepackage{xcolor}
\usepackage{placeins}

\begin{document}

\title{Entanglement-based quantum digital signatures over deployed campus network}

\author{Joseph C. Chapman,\authormark{1,*} Muneer Alshowkan,\authormark{1} Bing Qi,\authormark{1,2} and Nicholas A. Peters\authormark{1}}

\address{\authormark{1} Quantum Information Science Section, Oak Ridge National Laboratory, Oak Ridge, TN 37831, USA\\
\authormark{2} Currently at NYU Shanghai, 567 West Yangsi Road, Shanghai 200126, China}

\email{\authormark{*}chapmanjc@ornl.gov} 


\begin{abstract*} 
The quantum digital signature protocol offers a replacement for most aspects of public-key digital signatures ubiquitous in today's digital world. A major advantage of a quantum-digital-signatures protocol is that it can have information-theoretic security, whereas public-key cryptography cannot. Here we demonstrate and characterize hardware to implement entanglement-based quantum digital signatures over our campus network. Over 25 hours, we collect measurements on our campus network, where we measure sufficiently low quantum bit error rates (<5\% in most cases) which in principle enable  quantum digital signatures at over 50 km as shown through rigorous simulation accompanied by a noise model developed specifically for our implementation. These results show quantum digital signatures can be successfully employed over deployed fiber. Moreover, our reported method provides great flexibility in the number of users, but with reduced entanglement rate per user. Finally, while the current implementation of our entanglement-based approach has a low signature rate, feasible upgrades would significantly increase the signature rate.
\end{abstract*}


\section{Introduction}
A digital signature is one of the most widely used cryptographic primitives offering authenticity and integrity, non-repudiation, and transferability of digital content~\cite{10.1007/978-3-319-93387-0_8}. Classical digital signature protocols are based on public-key cryptography, such as RSA~\cite{10.1145/359340.359342}, which cannot provide information-theoretic security. To achieve information-theoretic security, two solutions have been developed recently, namely, unconditional secure signature (USS)~\cite{10.1007/978-3-319-93387-0_8} and quantum digital signature (QDS)~\cite{gottesman2001quantum,PhysRevA.91.042304,PhysRevA.93.032316}. In USS, pair-wise secret keys can be generated through quantum key distribution (QKD) protocols, and then the generated keys are employed to implement a secure digital signature. In contrast, QDS can be implemented with correlated (but not perfectly identical) raw keys, and can tolerate more channel loss. Both USS and QDS protocols can be implemented using the same quantum communications infrastructure for QKD, thus can extend the functionalities of a QKD network. However, both require pairwise quantum communications. Due to this restriction, they do not offer the universal verifiability property inherent to standard public-key digital signature schemes and are not a suitable replacement for many core applications of digital signatures~\cite{10.1007/978-3-319-93387-0_8}. Nevertheless, they might be useful in the cases when long-term, strong security is required. Moreover, the transferability of the USS and QDS protocols could have applications, such as, signed voting or enable creation of a centralized certification authority~\cite{Pelet_2022}.

QDS protocols have been demonstrated using different QKD systems, including decoy state BB84 QKD~\cite{An:19}, measurement-device-independent (MDI) QKD~\cite{PhysRevA.94.022328,Roberts2017,PhysRevA.95.042338,Collins2017}, and continuous-variable (CV) QKD~\cite{PhysRevLett.117.100503,PhysRevA.99.032341}. QDS has also been demonstrated using polarization entanglement, where users are provisioned different bandwidths of the entanglement spectrum~\cite{Pelet_2022}.

In this work, we develop an entanglement-based QDS protocol and demonstrate hardware for it on a deployed campus network. We use post-selected entanglement of a degenerate type-II pair source which enables a beamsplitter tree to simply provision entanglement between as many users as the beamsplitter tree has outputs. For this demonstration, we develop an automated QBER calibration system leveraging our control plane~\cite{PRXQuantum.2.040304} to transmit synchronized time tags and to control equipment across the network. Moreover, we employ simulations of our QDS protocol to show QDS performance using the measurements of our developed hardware as well as with various potential upgrades.

Our work has similarities and differences from previous QDS demonstrations in several aspects. Similar to Ref.~\cite{PhysRevA.95.042338,Collins2017}, our work is also demonstrated over deployed optical fiber. Ref.~\cite{Collins2017} co-locates Bob and Charlie using the same equipment, whereas in Ref.~\cite{PhysRevA.95.042338} and our work, Alice, Bob, and Charlie are in different physical locations connected by deployed optical fiber. Unlike Ref.~\cite{An:19,PhysRevA.94.022328,Roberts2017,PhysRevA.95.042338,Collins2017,PhysRevLett.117.100503,PhysRevA.99.032341} but similar to Ref.~\cite{Pelet_2022}, our work uses entanglement, specifically polarization entanglement. Our work differs from Ref.~\cite{Pelet_2022} in that each users in our demonstration is provisioned polarization entangled photons of the same wavelength, whereas, Ref.~\cite{Pelet_2022} uses wavelength multiplexing to provision entanglement to different users.

Finally, a QDS protocol (or more generally, quantum cryptographic protocols) could be implemented using either entanglement-based schemes (as we have done and Ref.~\cite{Pelet_2022}) or prepare-and-measure schemes~\cite{An:19,PhysRevA.94.022328,Roberts2017,PhysRevA.95.042338,Collins2017,PhysRevLett.117.100503,PhysRevA.99.032341}. While many prepare-and-measure schemes are implementation-friendly by employing widely available laser sources, entanglement-based protocols have their own advantages. More specifically, in our implementation, the intrinsic randomness of the entangled source is harnessed. There is no need for performing random number generation, active encoding, or phase randomization. Furthermore, our design is based on a commercially available off-the-shelf entangled-photon source, implying a path for potential commercialization. Another benefit of entanglement-based protocols is that the photon source can be untrusted. This removes potential side-channel attacks against the source and thus improve the practical security of the whole system~\cite{shannon2020use} but, when using spontaneous parametric processes, this comes with the trade-off of lower key rates due to attempts to suppress multi-pair emission by weak pumping of the entangled photon source. 

\section{Protocol and Security Analysis}
\label{sec:secanalysis}
\subsection{Quantum digital signature protocol}
We adopted the QDS protocol developed in  Ref.~\cite{PhysRevA.93.032325,Collins2017}. In the case of three-party (Alice, Bob, and Charlie) QDS, the protocol can be summarized as follows:

Distribution stage:
\begin{enumerate}
    \item For each possible future message $m = 0$ or 1, Alice generates raw keys $A_m^B$ and $A_m^C$ with Bob and Charlie through insecure quantum channels and authenticated classical channels. As in the case of QKD, an authenticated channel is required to prevent man-in-the-middle attacks. The corresponding raw keys held by Bob and Charlie are $K_m^B$ and $K_m^C$. We assume the length of each raw key is $L$. Alice’s signature for the future message $m$ is $Sig_m=(A_m^B,A_m^C)$.
    \item Bob and Charlie symmetrize their raw keys by randomly choosing and exchanging half of the bit values of $K_m^B$ and $K_m^C$ using a secure classical channel. The generated symmetric keys are $S_m^B$ and $S_m^C$. We remark that the above secure classical channel could be established by first conducting QKD between Bob and Charlie and then encrypting the raw keys to be exchanged (which are treated as data) by using one-time-pad. This process may consume a large amount of QKD key ($L$ bits for $K_0^{B.C}$ and $K_1^{B,C}$). To be more efficient, we could move this symmetrization process to the Messaging Stage (after Bob or Charlie receives Alice's message and the corresponding signature). In this case, Bob and Charlie may only need an authenticated classical channel rather than a secure classical channel. We will not consider the cost of symmetrization in this paper.
    
\end{enumerate}

More specifically, to generate the raw key in our implementation (see details in Sec.~\ref{sec:methods}), the output of a co-linear degenerate type-II spontaneous parametric down-conversion (SPDC) source is divided into three spatio-temporal modes and distributed to Alice, Bob, and Charlie, respectively. Each recipient performs a polarization measurement in a randomly chosen basis (X or Z) and records the detection result, the basis information, and the detection time. Through a classical authenticated channel, any pair of users can post-select the detection events when they use the same measurement basis, and the detection times are within a predetermined coincident window. The corresponding detection results form the raw key between the two users which is then sifted to only include events where the same basis was measured by both parties.

Messaging Stage:
\begin{enumerate}
    \item To sign one-bit message $m$, Alice sends $(m, Sig_m)$ to the desired recipient (say Bob).
    \item Bob counts the mismatches between $Sig_m$ and the two halves of $S_m^B$ (the half of $K_m^B$ which is not shared with Charlie and the half of $K_m^C$ received from Charlie) separately. If both are less than $s_a\frac{L}{2}$, Bob accepts the message. Here $s_a<\frac{1}{2}$ is a predetermined constant.
    \item To forward the message to Charlie, Bob forwards the pair $(m, Sig_m)$.
    \item Charlie counts the mismatches between $Sig_m$ and the two halves of $S_m^C$ separately. If both are less than $s_v\frac{L}{2}$, Charlie accepts the forwarded message. Here $0<s_a<s_v<\frac{1}{2}$. Note that it is a general feature of secure digital signatures that the thresholds for direct message and forwarded message are different \cite{swanson2011unconditionally}.
    
\end{enumerate}

To evaluate the performance of QDS, we consider the following properties: the robustness, the security against repudiation, and the security against forging. According to Ref.~\cite{Collins2017}, the probability of protocol failure under honest conditions is given by
\begin{equation}
    P(\text{Honest Abort})\le2e^{-{(s_a-E)}^2L}.\label{eq:PAbort}
\end{equation}
where $E$ is the observed bit error rate of the raw keys.

The probability of repudiation (the probability that Alice generates a signature which is accepted by Bob, but rejected by Charlie when forwarded) is given by
\begin{equation}
    P(\text{Repudiation})\le2e^{-{(\frac{s_a-s_v}{2})}^2L}
\end{equation}
and the probability of forging (the probability that Bob or Charlie accepts a signature not from Alice) is given by
\begin{equation}
    P(\text{Forge})\le2e^{-{(P_e-s_a)}^2L}.
\end{equation}
Here $P_e$ is the minimum error rate an adversary can achieve when guessing the honest party's raw key. In general, $P_e$ can be determined from the error rate $E$ of the raw key and the properties of the underlying QKD protocol. Note that the parameters $S_a$ and $S_v$ are crucial when the lengths of the raw keys are finite and should be optimized based on system parameters (see below). Generally speaking, from the experimentally observed quantum bit error rate $E$, Alice and Bob can determine the minimum error rate $P_e$ that an adversary can achieve when guessing the honest party’s raw key. Obviously, a necessary condition for security is $E < P_e$. In the asymptotic case (meaning the raw key length is infinite), $S_a$ could be set to any value between $E$ and $P_e$ without compromising the robustness of the protocol or the security against forging. However, when the raw key length is finite, we need to consider statistical fluctuations and make a trade-off between robustness and security against forging by choosing an appropriate $S_a$ value. Similar arguments apply to $S_v$.

To generate the raw keys in our implementation, we assume Alice shares entangled photon pairs with Bob and Charlie and each user randomly selects between measuring the received photons in the Pauli X or Pauli Z basis. In this entanglement-based protocol, secret key rate is given by \cite{koashi2003secure}:
\begin{equation}
    R \ge Q (1-f H_2(E_Z)-H_2(E_X)),
\end{equation}
where $f$ quantifies the efficiency of the QKD error correcting algorithm.

Since there is no need to do error correction in the QDS protocol, secure QDS is possible as long as $1-H_2(E_Z)-H_2(E_X)>0$. This suggests that $P_e$ can be determined from $H_2(P_e)=1-H_2(E)$. Here we assume measurements in the two bases are sampled evenly and are combined; so the gain is $Q=Q_Z+Q_X$ and quantum bit error rate (QBER) is $E=\frac{E_X+E_Z}{2}$. 

The QDS protocol is $\epsilon$ secure if
$\text{max}[P(\text{Forge}), P(\text{Repudiation}), P(\text{Honest Abort})]\leq\epsilon$. In practice, it is reasonable to choose $P(\text{Forge})=P(\text{Repudiation})=P(\text{Honest Abort})$. Under this condition, from Eqs.(1)-(3), we have
\begin{equation}
    s_a=E+\frac{P_e-E}{4}
\end{equation}
\begin{equation}
    s_v=E+\frac{3(P_e-E)}{4},
\end{equation}

In the finite data cases, given the desired security levels and QDS system parameters, we can determine the required raw key length ($L$)  using the above equations. Correspondingly, given the transmission rate ($v$) of quantum communications, we can determine the number of signature keys that can be generated per second as
\begin{equation}
R_{QDS}=vQ/L.
\end{equation}

\subsection{Noise model for entanglement-based QDS}

\begin{figure}
\centerline{\includegraphics[width=0.75\textwidth]{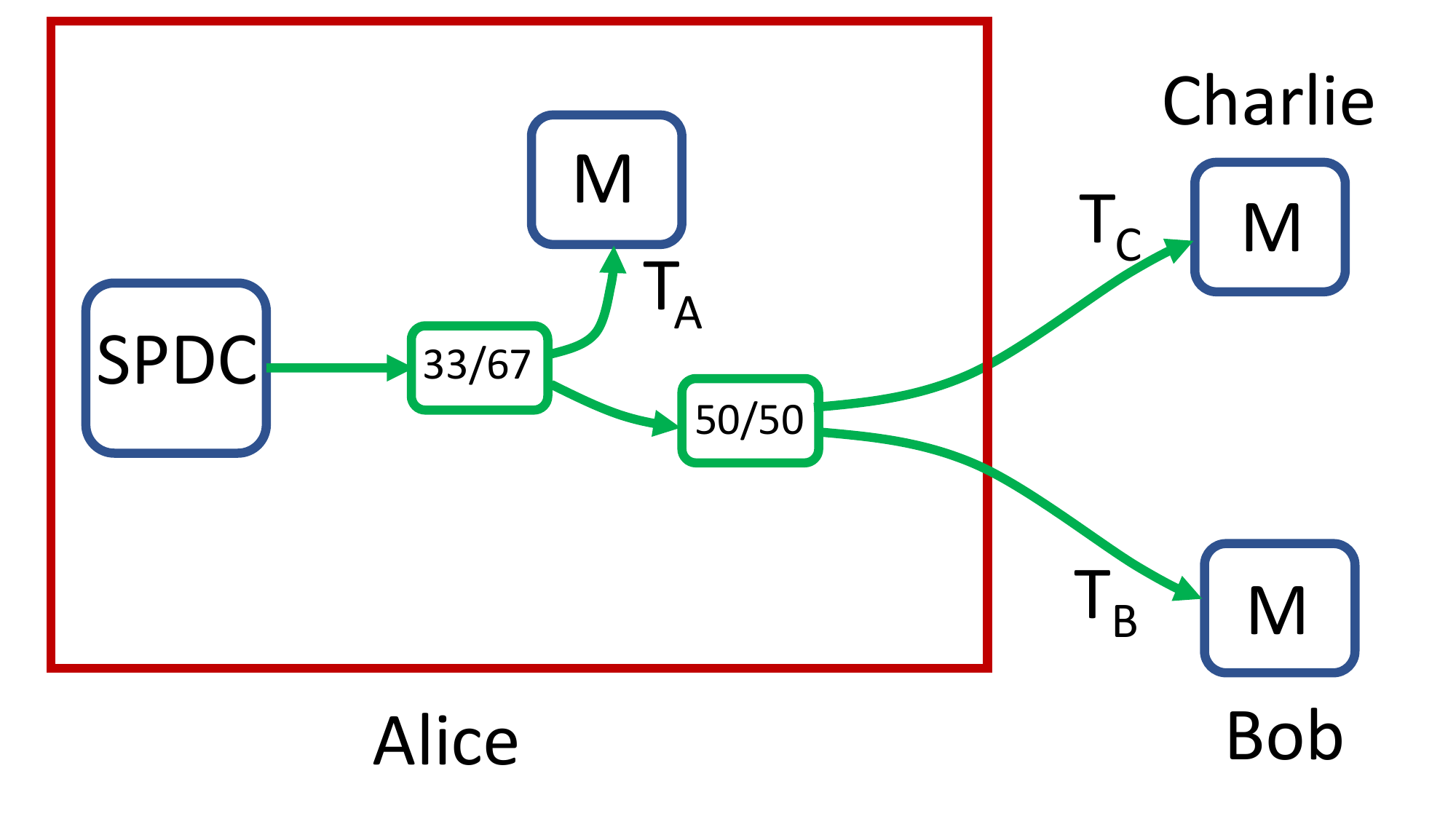}}
\caption{Schematic of three-party quantum digital signatures using our noise model. Polarization entanglement is generated by Alice using spontaneous parametric downconversion (SPDC) and is split to three parties (Alice, Bob, and Charlie) using beamsplitters and transmitted to their measurement systems where they each make a polarization measurement in one of two mutually unbiased bases.}
\label{fig:noisemodel}
\end{figure}

While the security of an entanglement based QDS protocol does not rely on an accurate modelling of the entangled source, such a noise model is very useful for parameter optimization and performance estimation. Here we develop a noise model for this purpose and apply it to the configuration in Fig.~\ref{fig:noisemodel}.

In this work, a commercial continuous-wave (CW) pumped, co-linear degenerate type-II spontaneous parametric down-conversion (SPDC) source is employed. Under the designed working conditions, the output of the source can be described by Ref.~\cite{TAKESUE2010276}
\begin{equation}
    \ket{\Psi_{dist,n}}=\prod_{i=1}^{n}\ket{1,1}_{HV,i},
\end{equation}
where $\ket{1,1}_{HV}$ represents a quantum state with 1 $H$-polarization photon and 1 $V$-polarization photon, and the subscript ``dist'' represents that different photon pairs are distinguishable. 

The probability of generating $n$-pair photons is given by Ref.~\cite{TAKESUE2010276}
\begin{equation}
    P(n)\cong e^{-2\lambda}\frac{{(2\lambda)}^n}{n!},\label{Pnpoisson}
\end{equation}
where $\lambda$ is half of the average photon pair number and can be adjusted experimentally to maximize secure key rate. This approximation is valid for $N=\Delta{t_{pump}}/\Delta{t_{SPDC}}>>n$, where $\Delta{t_{pump}}$ is the pump pulse width and $\Delta{t_{SPDC}}$ is the coherence time of the photon pair. We assume the SPDC source is ideal except with an internal loss of $\eta_0$. Any polarization imperfection at the source could be absorbed into the polarization misalignment of the measurement system. We further assume that the SPDC source is weakly pumped $\lambda<<1$.

The polarization measurement system is based on the passive basis-selection scheme using four identical single-photon detectors (SPDs). The passive basis selection is implemented using a 50/50 beam splitter.
Given four SPDs in each polarization measurement system, there are in total $2^4=16$ possible distinct detection events. The post-selection is based on the following rules~\cite{PhysRevLett.101.093601,PhysRevA.89.012325}: (1) all no-click events are thrown away; (2) all events with cross-detection (SPDs in different bases fire simultaneously) are discarded; (3) a double-click event in the same basis is assigned a random bit value; (4) all single counts are kept.

We assume that the probability of more than one dark count within one detection window is negligible. All the detectors for a given user are identical (the same performance) with detection efficiency of $\eta_{D,i}$ (where $i\in \{1,2,3\}$ for Alice, Bob, and Charlie, respectively), which includes the measurement system efficiency and the basis-choice probability, and dark-count probability of $D_0$ where the higher dark-count probability is used for all detectors if they differ between users. Errors due to polarization misalignment are quantified by $e_{dx}$ ($X$-basis errors) and $e_{dz}$ ($Z$-basis errors), i.e., $e_{dz}$ is the probability that an $H$-photon is registered by the $V$-detector, or vice versa.

To derive the gain and QBER, we analyze the lower order terms individually, then analyze the higher order ones as a group. The gain $Q$ and QBER $E$ are defined as
\begin{equation}
    Q=\sum^{\infty}_{n=0}{P(n)Y_n},\text{ and}
\end{equation}
\begin{equation}
    EQ=\sum^{\infty}_{n=0}{e_nP(n)Y_n},
\end{equation}
where $Y_n$ is the yield. We will also make use of $T_A=\eta_0 t_1\eta_D$, $T_j=\frac{\eta_{0}(1-t_1)\eta_c\eta_D}{2}$, where $j$ denotes $B$ (Bob) or $C$ (Charlie), $\eta_c$ is the channel loss, $t_1$ is the beamsplitter transmission from the source to Alice assuming $1-t_1$ is directed toward a 50/50 beamsplitter between Bob and Charlie (See Fig.1 for the experimental setup). To analyze the $Y_n$ terms we look at the different photon numbers emitted by the source as detected in the $X$ and $Z$ bases, we also look at the terms in the QBER at the same time; details are presented in ~\ref{App:NM}.


\section{Experimental Methods}
\label{sec:methods}
\begin{figure}
\centerline{\includegraphics[width=1\textwidth]{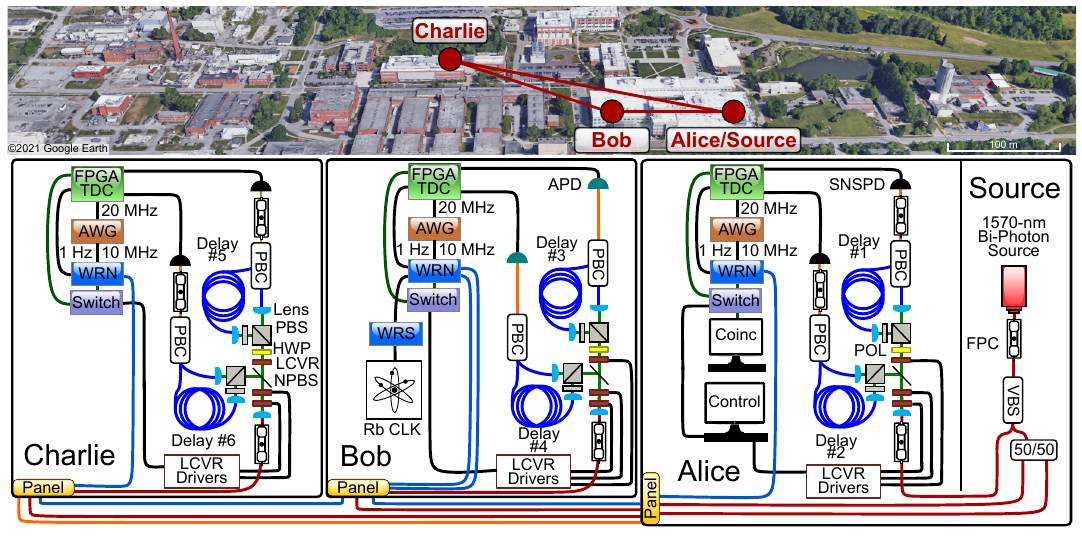}}
\caption{Quantum digital signatures source and receiver setup. Alice's source produces (post-selected) entangled photon pairs which are split between three users: Alice, Bob, and Charlie. Each user has a receiver to measure the received photons in the horizontal/vertical or diagonal/anti-diagonal polarization bases. The two output modes of each basis measurement are time-multiplexed using a delay fiber and polarization beam combiner then directed to single-photon detectors whose outputs are timetagged for analysis. Delays 1-6 ranged from 1 m to 7 m and were all different to ensure the time-multiplexed signals could be distinguished in the coincidence analysis. APD = avalanche photodiode. AWG = arbitrary-waveform generator. Coinc = Coincidence analysis. FPC = fiber polarization controller. FPGA TDC = field-programmable-gate-array-based time-to-digital converter. HWP = half-wave plate. LCVR = liquid-crystal variable retarder. NPBS = non-polarizing beamsplitter. PBC = fiber-based polarization beam combiner. PBS = polarizing beamsplitter. Rb CLK = Rubidium atomic clock. SNSPD = superconducting-nanowire single-photon detector. WRN = White-Rabbit Node. WRS = White-Rabbit Switch.}
\label{fig:setup}
\end{figure}

\begin{figure}
\centerline{\includegraphics[width=0.75\textwidth]{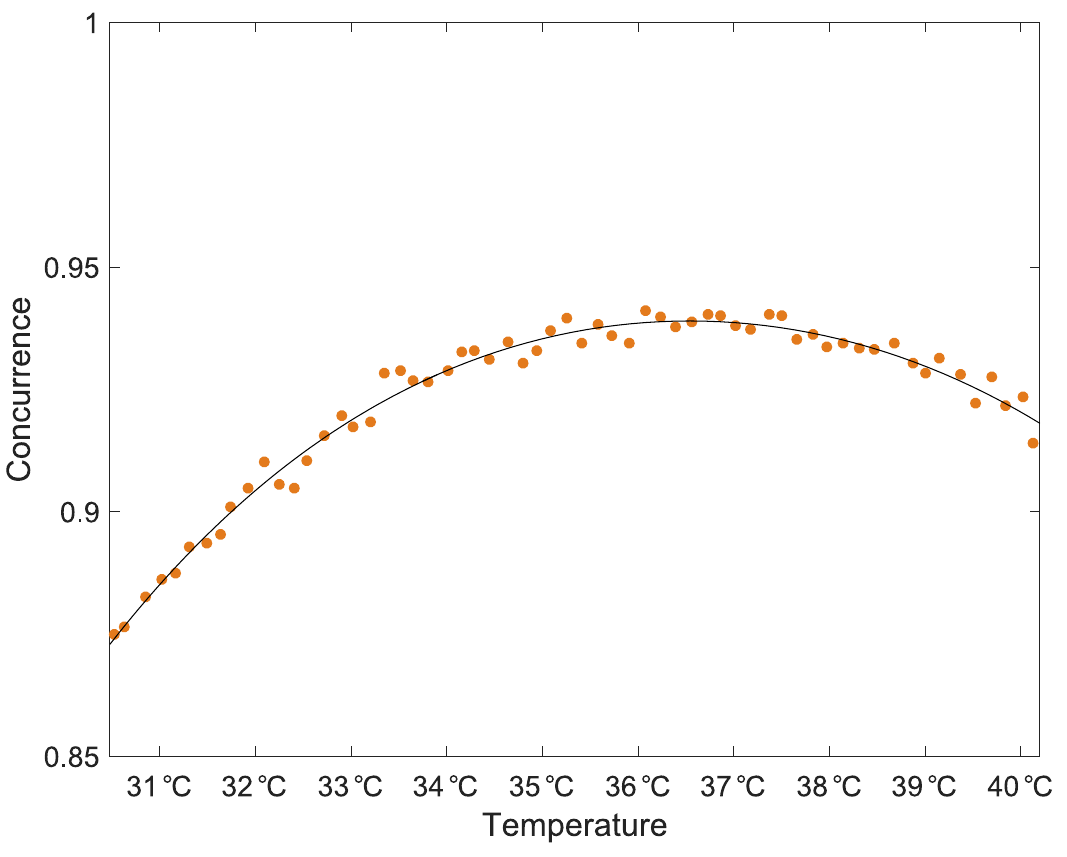}}
\caption{Source entanglement quality characterization using crystal temperature sweep. A peak concurrence of 0.94 is found at 36.5$^{\circ}$C with coefficient of determination R$^2$=0.999993.}
\label{fig:QEStempopt}
\end{figure}

Our quantum digital signatures implementation (Fig.~\ref{fig:setup}) starts with a two-photon quantum entanglement source (QES, Qubitekk, Inc.) emitting Horizontal-Vertical photon pairs centered at 1570.2 nm with about 1.5-nm full-width-at-half-maximum bandwidth. When the spectral distribution of the horizontal and vertical photons overlaps perfectly, this source generates two-mode polarization entanglement conditioned on a photon detection in each mode, i.e, post-selected entanglement. The entanglement is optimized by sweeping the crystal temperature then (with a separate setup~\cite{PhysRevApplied.19.044026}) measuring a two-mode polarization quantum state tomography~\cite{altepeter2005photonic} using maximum likelihood analysis~\cite{PhysRevA.64.052312} to calculate the concurrence~\cite{PhysRevLett.78.5022} from the reconstructed density matrix. Fig.~\ref{fig:QEStempopt} shows the concurrence for a range of crystal temperatures. The curve is fit~\cite{nlmfit} to a 4th-order polynomial to find the peak location. The peak concurrence (about 0.94) is found to be at 36.5$^{\circ}$C; more importantly, since the crystal-temperature sensor is itself noisy ($\pm0.1^{\circ}$C), the peak is found when setting the temperature controller digital-to-analog converter to 30200.

To preserve the polarization entanglement, the photons are separated into different modes using beamsplitter(s). This provides a very flexible method for provisioning this post-selected entanglement to a variety of users. Here a beamsplitter tree is configured to provide entanglement between three pairs of users: Alice-Bob, Alice-Charlie, and Bob-Charlie. An evanescent variable fiber beamsplitter (Newport F-CPL-1550-N-FA) is calibrated to send about 1/3 to Alice and 2/3 to a 50/50 fused coupler (Thorlabs TN1550R5F2) so that Bob and Charlie also end up with 1/3 each. The transmission from the source to the 3 outputs of the beamsplitter tree is 0.77 (when summing all outputs).

After the beamsplitter tree, Alice's mode is directed into her analyzer, whereas Bob's and Charlie's modes propagate about 250 m (2.44-dB loss) and 1200 m (5.5-dB loss) through the campus fibers to their analyzers, respectively. Upon entering the analyzer, the photons pass through a manual polarization controller and two liquid-crystal variable retarders (LCVR, Thorlabs LC1113-C and LC1513-C), rotated about 45$^{\circ}$ with respect to one another, for automated polarization control. Then the photons propagate to a 50/50 cube non-polarizing beamsplitter (NPBS, Thorlabs CCM1-BS015) to implement a passive basis choice between the horizontal/vertical (H/V) and diagonal/anti-diagonal (D/A) bases. The photons reflected by the 50/50 beamsplitter are measured in the H/V basis primarily using a cube polarizing beamsplitter (PBS, CCM1-PBS254); but due to imperfect reflection-port extinction ratio, a vertically oriented clean-up polarizer (Thorlabs LPNIRC050-MP2) is placed in the reflection port. This clean-up polarizer is crucial to guaranteeing a low QBER because the PBS reflection-port extinction ratio is often $<<$1:100. Similarly, the D/A basis measurement uses a PBS and clean-up polarizer but in front of the PBS is a half-wave plate (HWP) (to rotate the incoming photons from the D/A basis into the H/V basis of the PBS) and another LCVR (with the slow axis in the H/V basis) to adjust the measurement phase for low D/A QBER. The output of each measurement port is collected into polarization-maintaining fiber with the slow axis oriented to match the photon polarization. To calibrate the axes of the polarization components with respect to one another, we used back-propagating classical light and an additional calibration polarizer (to make crossed polarizers with the PBS for the HWP and LCVR). After alignment and calibration, the total receiver optical transmission (when summing all output ports) is 0.37, 0.225, and 0.3175 for Alice, Bob, and Charlie, respectively.

After collecting the receiver outputs into a polarization-maintaining fiber, we time-multiplex the two output modes from each basis measurement using a delay fiber and a polarization beam combiner (PBC). Each delay fiber was a different length, ranging from 1 to 7 m, to ensure the time-multiplexed signals could be distinguished in the coincidence analysis. We then direct the single-mode port of the PBC towards a single-photon detector. Therefore, this setup only requires each receiver to use two single-photon detectors: one for the H/V basis and another for the D/A basis. This time-multiplexing is done due to practical constraints on the number of suitable detectors available and has little-to-no impact on our results because the overall count rates are low enough to have very low probability of multi-photon detection due to time-multiplexing. With more detectors, it is straightforward to remove the delay fiber and polarization-beam combiner and replace the delay with another detector for each basis.
While Alice and Charlie utilize SNSPDs, each preceded by an fiber polarization controller to balance the detection efficiency on every basis, Bob uses APDs (IDQ Qube) free-running set at an efficiency of 20\% and deadtime of 10 $\mu s$.

The output pulses of the detectors in each location are timetagged by a field-programmable gate array (FPGA) time-to-digital converter (TDC) which is time synchronized by the White Rabbit timing synchronization system. In Bob's location, a Rubidium (Rb) atomic clock (Stanford Research Systems FS725) supplies a 10-MHz clock for a White Rabbit switch (WRS) through a direct coaxial connection. This WRS connects to each White Rabbit node (WRN) using an optical fiber. The 10-MHz output from each WRN is then doubled using an arbitrary waveform generator (AWG), then feeds the FPGA-based TDC to synchronize their time-stamping clocks~\cite{PRXQuantum.2.040304,Alshowkan:22}. Additionally, the White Rabbit system provides a pulse-per-second signal for the TDC. Besides its remarkable timing synchronization capabilities, the White Rabbit system also offers ethernet-based networking that we use for classical communications. Thus, the timing synchronization and classical communications are carried over a single optical fiber. 

The coincidence analysis server (CAS) initiates a measurement process by calling all FPGA-based TDCs to timestamp the detectors' output events and subsequently stream this data back where it is stored in distinct files. Following each measurement cycle, the CAS performs a cross-correlation of file pairs, using an offset to account for the optical signal detection delay at all locations.
To optimize performance, we employ a binary file format for both writing and reading timestamp files while harnessing the power of a graphics processing unit (Nvidia Quadro P6000) to parallelize the 
cross-correlation task efficiently.
This process results in a coincidence histogram. By applying a predefined delay offset and a coincidence window of approximately 2 ns, coincidence events are extracted for each unique measurement setting. Finally, the QBER is computed from these coincidences and the result is transmitted to the control machine for calibration.

This detection and analysis system requires calibration of several experimental parameters, e.g., photon pairs from the source will travel different pathlengths to the various detectors. To accurately identify coincidence counts between the detectors, we calibrated the delays between all of the detectors for the cases where each photon goes to a different user. This calibration is done by looking at the cross-correlation of the timetags between a pair of detectors while only transmitting one of the time-multiplexed inputs. A single correlation peak is found at the optimal delay. While it is feasible to fine-tune a small delay window for each pair of detectors, adopting a uniform delay window makes it easier to troubleshoot, ensuring consistency and reducing overall complexity.

Moreover, with these coincidences (used to calculate the QBERs), the receiver LCVRs need to be calibrated to minimize the measured QBERs between each pair of users. This is not a trivial optimization task due to the 8-dimensional solution space, noisy QBER measurements, and low sample rate due to long integration times. To provide a good starting point for QBER optimization using LCVRs, we first approximately calibrate the H/V QBER using a classical laser. We connect a 1565-nm laser to the output PM fiber of the source and adjust the manual polarization controller at each receiver to minimize the light detected in the V-output of each receiver using a calibrated power meter (Thorlabs S120C and PM400). Due to various imperfections in the system, this only minimized each H/V QBER to about 20\%. 

To further calibrate the H/V QBERs and then calibrate the D/A QBERs between all three receivers across our network, we develop and employ a multi-state cyclic coordinate descent~\cite{Vassiliadis2009} algorithm, which leverages QBER optimization by curve-fitting of algorithmically determined measured samples and numerical optimization of the resulting fitted curve; algorithm details are in ~\ref{app:optdetails}. 
After optimizing the QBER using our multi-state cyclic coordinate descent algorithm, all H/V and D/A QBERs between all three pairs of users should be minimized. The hardware is now ready for data acquisition and key generation.

\section{Results}
\label{sec:results}
For quantum cryptography hardware, the QBER and sifted raw key rate are two major characteristics of a given implementation. In addition, specific to quantum digital signatures, the sifted raw key length per signature and signature rate are also of interest. Using measurement and simulation, we investigate these and other interesting characteristics of our implementation.

After the aforementioned QBER calibration, to characterize our system we collected ten datasets using a 30-s integration time for each dataset. Across our campus network, the average sifted raw key rate between Alice-Bob, Alice-Charlie, and Bob-Charlie is 71$\pm2$ bits/sec, 88$\pm3$ bits/sec, and 21$\pm1$ bits/sec, respectively. The average QBER for Alice-Bob H/V basis, Alice-Charlie H/V basis, Bob-Charlie H/V basis is 0.043$\pm0.004$, 0.034$\pm0.005$, and 0.038$\pm0.008$, respectively. Similarly, the average QBER for Alice-Bob D/A basis, Alice-Charlie D/A basis, Bob-Charlie D/A basis is 0.069$\pm0.006$, 0.046$\pm0.004$, and 0.047$\pm0.016$, respectively. The Alice-Bob D/A QBER is somewhat higher than the rest due to some residual phase error from the calibration process. The QBER calibration process shows the ability to bring all QBERs low enough for quantum cryptography protocols but does not necessarily converge perfectly due to noisy photon counting data. The specific impact of these values and possible improvements will be evaluated below in simulation but overall these values are low enough to execute QKD and QDS using finite-key analysis.

\begin{figure}
\centerline{\includegraphics[width=1\textwidth]{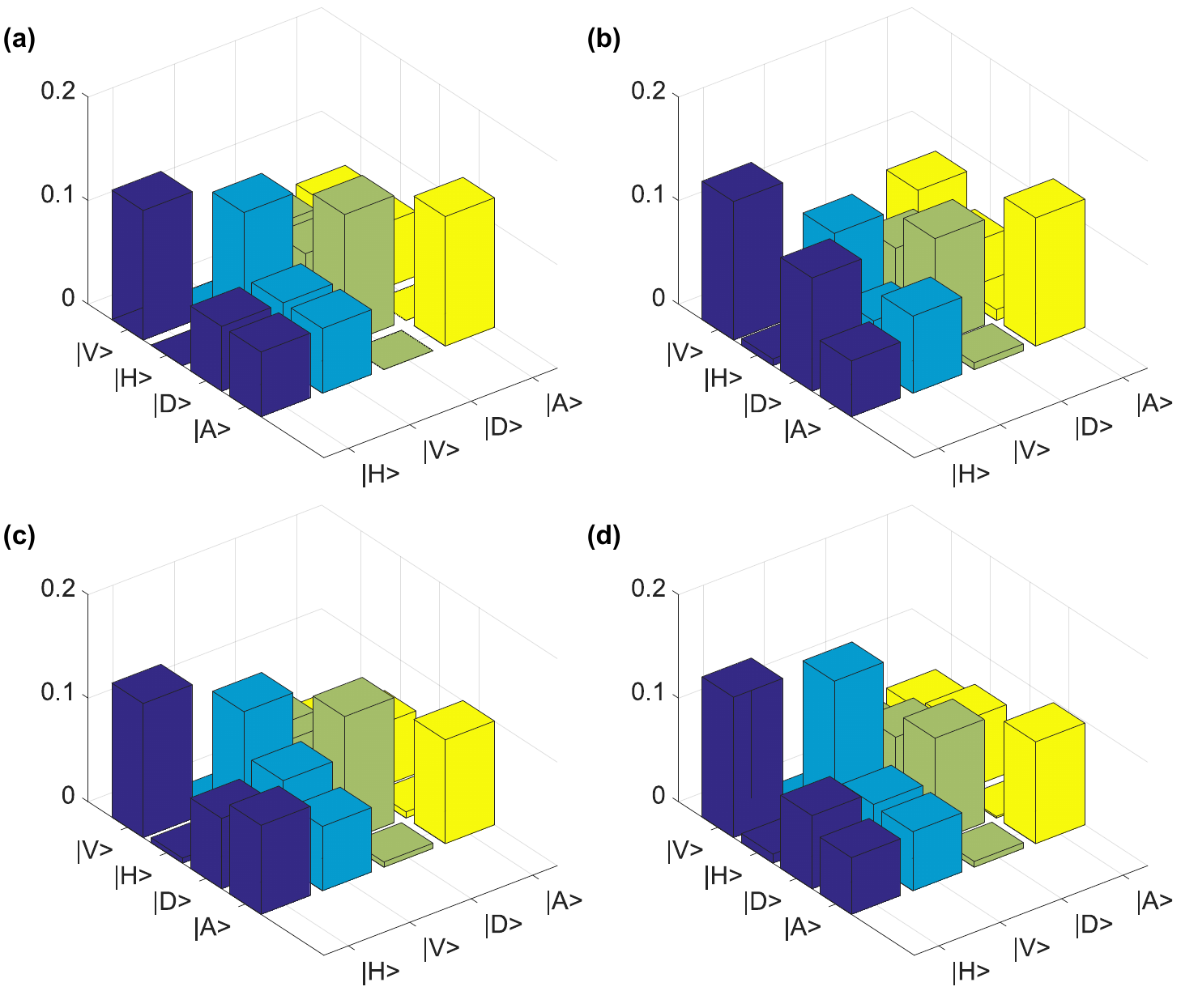}}
\caption{Normalized crosstalk matrices. These matrices are each normalized to the sum of all matrix elements, i.e., all coincidence counts between that pair of users. Alice's (Bob's) projective measurements is shown on the y-axis (x-axis). (a) Ideal (b) Alice-Bob. (c) Alice-Charlie. (d) Bob-Charlie.}
\label{fig:crosstalk}
\end{figure}

The coincidences between a pair of users can be visualized in matrix form as a crosstalk matrix (Fig.~\ref{fig:crosstalk}). Ideally [Fig.~\ref{fig:crosstalk}(a)], when each user measures in the same basis the results should be strongly correlated (i.e., the 2x2 submatrix is diagonal). When they measure in different bases there is ideally no correlation (i.e., the 2x2 submatrix is uniform). The ratio in probability between upper diagonal submatrix (both users measuring in H/V basis) and the lower diagonal submatrix (both users measuring in D/A basis) provides a measurement of the effective passive basis choice probability (effective because it includes the relative efficiency of the measurements). For any given basis choice, the QBER can be calculated by the ratio of the off-diagonals of that 2x2 submatrix to the total of all 4 elements. Due to our source emitting anti-correlated (H-V) pairs, one user in every pair flips their assignments of ``1'' and ``0'' (see axis labeling in Fig.~\ref{fig:crosstalk}). 

\begin{figure}
\centerline{\includegraphics[width=1\textwidth]{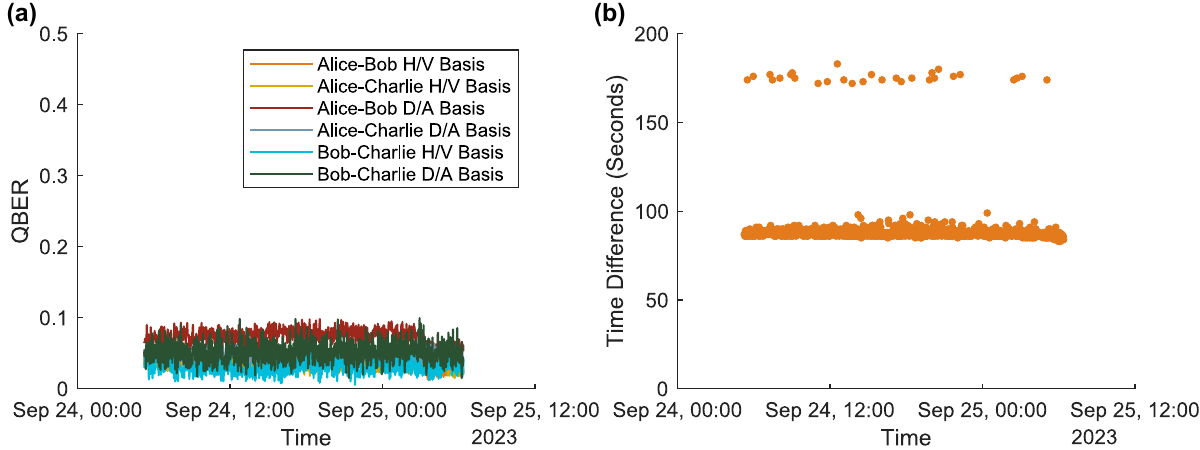}}
\caption{Extended QBER measurements. (a) QBER between all users over 25 hrs. Every 90 s, data is collected for 30 s and analyzed in real-time. There are 1003 saved measurements. (b) Time difference between successive datasets showing that some intervals were greater than 90 s.}
\label{fig:qbervtime}
\end{figure}

Moreover, after these ten datasets were collected we continued to collect measurements for about 25 hrs. The QBER for all datasets is plotted in Fig.~\ref{fig:qbervtime}(a). The average QBER for Alice-Bob H/V basis, Alice-Charlie H/V basis, Bob-Charlie H/V basis is 0.039$\pm0.008$, 0.033$\pm0.006$, and 0.03$\pm0.01$, respectively. Similarly, the average QBER for Alice-Bob D/A basis, Alice-Charlie D/A basis, Bob-Charlie D/A basis is 0.074$\pm0.01$, 0.045$\pm0.006$, and 0.052$\pm0.015$, respectively. Comparing the above statistics from the initial ten samples to these for 25 hrs, the QBER stayed relatively stable throughout---without active feedback and no pauses to recalibrate---as seen by the very similar averages and standard deviations. This is due in large part to indoor or underground fiber and no outdoor aerial fiber. 

Due to the automated nature of the QBER calibration, the QBER server was set up to begin saving data when all QBERS were below 10\% to ensure QDS data was saved as soon as it was likely viable; the QBER server was configured thus to automate the initiation of QDS data acquisition. The 10\% threshold is a more conservative implementation of the common QKD abort threshold of QBER >11\% implemented in our simulations. Over our 25-hr data collection, 3\% of the time (29/1003) times the measurements did not meet this criterion so they were incidentally not saved. Alice-Bob's D/A QBER average was 0.074; accordingly, there are likely some of those measurements which did not meet the threshold so the average quoted above for the 25-hr data is likely a little higher; but this did not affect the 10 consecutive datasets above. There is also another unrelated reason the QBERs would not have met this threshold: If the FPGA TDCs do not sync on the same White-Rabbit pulse-per-second signal then the coincidence delays are no longer correct and there are just accidental coincidences calculated with high QBER. This is likely to have been the cause for some fraction of the 3\% as this has been observed to occasionally happen during other measurements.

Now using this detailed characterization of our system, we simulate its performance using the analysis developed in Sec.~\ref{sec:secanalysis}, as if the post-processing of quantum digital signatures were fully implemented, and compare it with the performance of an upgraded system. The QDS protocol is dependent on the raw key rates of Alice-Bob and Alice-Charlie being equal. In reality, this is not always the case so the lower rate will determine the raw key rate. In addition, the higher QBER will determine the required raw key length per signature. Since our key rates and QBERs are different between different users, we calculate the QDS performance showing each possible pair as the limiting connection to aid the evaluation of the differences. Moreover, we also show the QDS metrics for the Bob-Charlie link for completeness, even though that link is not necessarily used to generate signatures. But this link could be used if the protocol roles are redistributed amongst the users, e.g., Alice and Charlie swap roles, but the equipment is not redistributed.

For the simulations, the polarization misalignment parameters $e_d$ need to be extracted from our QBER measurements (which also depend on detector noise and multi-photon errors). Using the simulated QBER,  we manually searched for the optimal $e_d$ that produced QBER matching our measurements. Our findings and the other relevant simulation parameters are displayed in Table~\ref{tab:simparams}; for which, several parameters depended on the methods in ~\ref{App:paramextract}.

\begin{table}
\centering
\caption{Simulation parameters for the current and what we might expect is possible in an improved system. $\eta_0$ = source internal efficiency. $\lambda$ = SPDC brightness parameter. $v$ = repetition rate (inverse coincidence window for CW pump). $\epsilon$ = security parameter. $D_0$ = detector background count probability per pulse. $\eta_D$ = combined measurement system transmission and detection efficiency. $e_d$ = polarization misalignment in the $X$-basis (D/A basis) and $Z$-basis (H/V basis).}
\label{tab:simparams}
\begin{tabular}{ccccccc}
\hline
Parameters & \multicolumn{3}{c}{Current} & \multicolumn{3}{c}{Improved}\\
\hline
$\eta_0$ & \multicolumn{3}{c}{0.3} & \multicolumn{3}{c}{0.5}\\
$\lambda$ & \multicolumn{3}{c}{0.0047} & \multicolumn{3}{c}{0.01}\\
$v$ & \multicolumn{3}{c}{$1/(2$ ns$)$} & \multicolumn{3}{c}{$1/(2$ ns$)$}\\
$\epsilon$ & \multicolumn{3}{c}{$10^{-10}$} & \multicolumn{3}{c}{$10^{-10}$}\\
\hline
&Alice&Bob&Charlie&\multicolumn{3}{c}{All users}\\
\hline
$D_0$ &$10^{-7}$& $10^{-6}$&$10^{-7}$ &\multicolumn{3}{c}{$10^{-8}$}\\
$\eta_D$  &0.14 &0.035 &0.12 &\multicolumn{3}{c}{0.324}\\
\hline
&Alice-Bob&Alice-Charlie&Bob-Charlie&\multicolumn{3}{c}{All user pairs}\\
\hline
$e_{dx}$  &0.031 &0.019 &0.018 &\multicolumn{3}{c}{0.01}\\
$e_{dz}$  &0.017 &0.013&0.013 &\multicolumn{3}{c}{0.01} \\
\hline
\end{tabular}
\end{table}

\begin{figure}
\centerline{\includegraphics[width=0.97\textwidth]{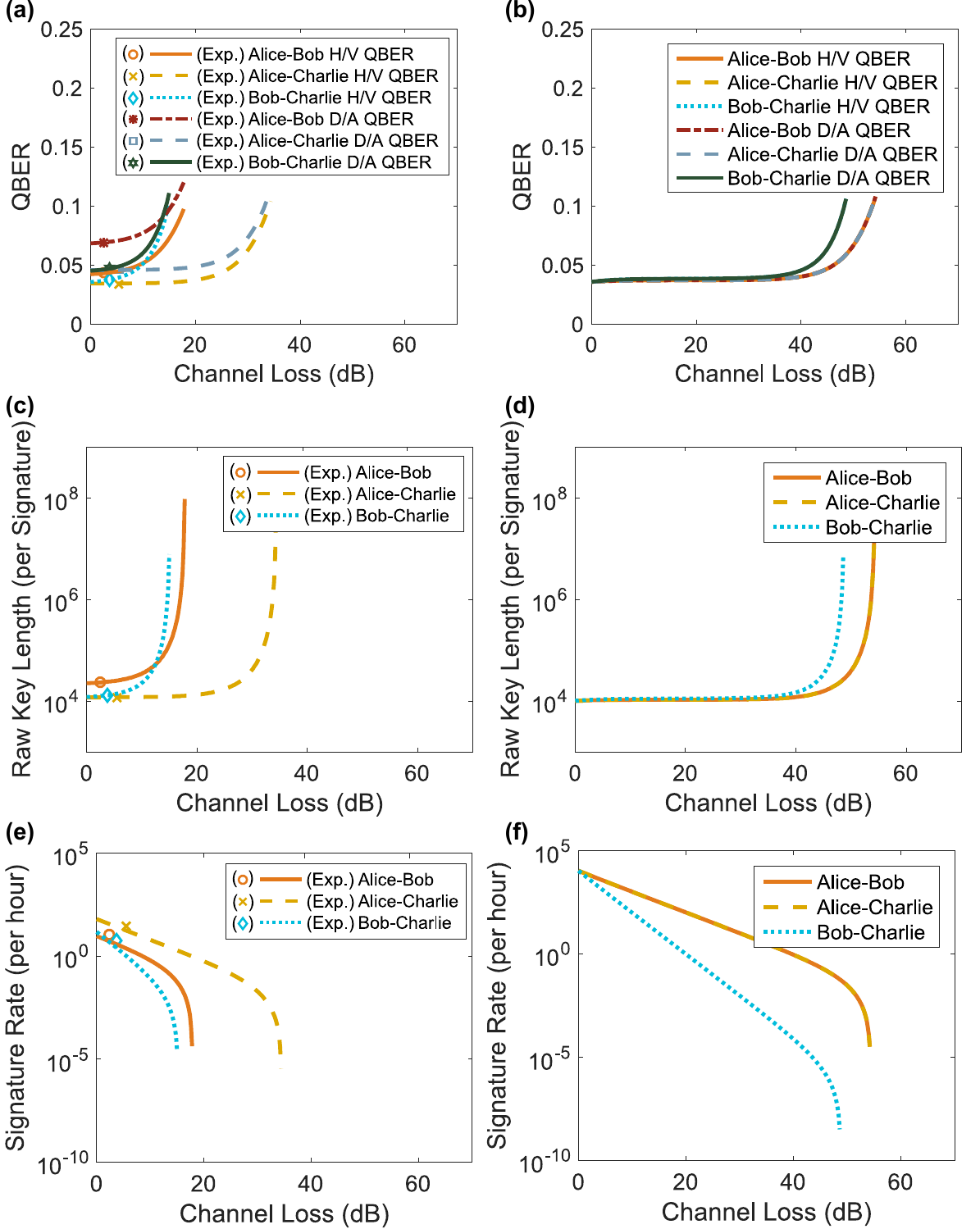}}
\caption{Quantum digital signatures for current and future improved system. Simulations using parameters from Table~\ref{tab:simparams}. (a) - (b) Measured and simulated QBER for each basis of all three user pairs for the current and improved (only in simulation) systems, respectively. (c) - (d) Experimentally estimated and simulated sifted raw key length (required per signature) versus channel loss of all three pairs of users for the current and improved (only in simulation) systems, respectively. For comparison, using low-loss fiber (0.2 dB/km), 20-dB channel loss corresponds to 100 km transmission distance. (e) - (f) Experimentally estimated and simulated signature rate versus channel loss of all three pairs of users for the current and improved (only in simulation) systems, respectively. Note, since Bob-Charlie data points experienced each channel's loss, we approximated the placement for this 2D plot by averaging their channel losses (2.4~dB and 5.5~dB).}
\label{fig:sims}
\end{figure}

With these simulation parameters derived from our current system and separate parameters which seem reasonable for an improved version, Fig.~\ref{fig:sims} displays QBER, (sifted) raw key length (required per signature), and signature rate (each versus channel loss). The channel loss refers to Bob and/or Charlie's channel loss since the source is with Alice, e.g., for Alice-Bob the channel loss is just Bob's but for Bob-Charlie it refers to both. In Fig.~\ref{fig:sims} (a), (c), and (e), the data points are estimated from our measurements. For Fig. ~\ref{fig:sims}(a) the data points are the QBER values reported at the beginning of Sec.~\ref{sec:results} and the channel losses reported in Sec.~\ref{sec:methods}. Notably, these QBERs were calculated using the entire 30-s data files and not just a portion as is done in finite-key QKD post-processing. For Fig. 5(c), the experimental required raw key length per signature data points are calculated using the measured QBERs (plotted in Fig.~\ref{fig:sims}(a)) and Eq.~(\ref{eq:PAbort}). For Fig.~\ref{fig:sims}(e), the estimated experimental signature rate is calculated using the measured raw key rates (reported at the beginning of Sec.~\ref{sec:results}) and the required raw key length per signature [plotted in Fig.~\ref{fig:sims}(c)]. Additionally, since our derived equations do not differentiate between the detector noise $D_0$ between the pair of users, we use the highest detector noise between the users to simulate all the detectors to conservatively simulate the worst-case scenario. Bob's APDs have higher noise than Alice or Charlie's SNSPDs which is in agreement with Fig.~\ref{fig:sims} showing the pairs including Bob do not tolerate as much channel loss as the Alice-Charlie pair. In general, these simulations validate that detector noise is a main limiting factor in maximum tolerable channel loss, in addition to source and measurement efficiency as well as uncorrelated multi-pair detection events, but more so than polarization misalignment; also found in similar work simulating different quantum cryptographic protocols~\cite{HEQKD}. In fact, if the polarization misalignments were not  improved from their current values, the improved system (with the rest of the improvements) could still generate signatures at over 50~dB based on our simulations; showing that low polarization misalignments play only a moderate role in the maximum operating loss.

Nonetheless, in Fig.~\ref{fig:sims}(c), difference in the raw key length per signature at low channel loss is dominated by the polarization misalignment as seen by Alice-Bob being the highest (Alice-Bob $e_{dx}$ is the highest of all). In Fig.~\ref{fig:sims}(e)-(f), the signature rate starts nearly the same for all users but drops off steepest for Bob-Charlie because both experience the channel loss.
With the current system's brightness and efficiency, our current implementation is only suitable for signing very small messages but still over significant distances; for example, using low-loss fiber (0.2 dB/km) even just 10-dB channel loss corresponds to 50 km transmission distance. Whereas, with modest improvements in noise, efficiency, and source brightness (chosen using optimization), the signature rate at low channel loss [Fig.~\ref{fig:sims}(f)] could increase by about 3 orders of magnitude and the maximum tolerable channel loss would increase by about 3 orders of magnitude leading to a maximum transmission distance in low-loss fiber of about 250 km (though at a very low signature rate). Clearly, this  protocol significantly benefits from low-channel-loss implementation. Interestingly, the raw key length per signature at low channel loss is nearly the same for the current and improved systems. This shows that the key length needed for a signature is more dependent on specifics of the protocol used than on the hardware performance.  Overall, our measurements and simulations show current technology is able to implement quantum digital signatures to sign short messages at low rates over moderate distances.

The improvements needed to obtain the performance shown in Fig.~\ref{fig:sims}(b), (d), and (f), could likely be achieved by integrating disparate pieces of current technology. To improve $\eta_0$, better mode-matching~\cite{PhysRevLett.115.250402,PhysRevLett.121.250505} or a waveguide-based entangled photon source could provide improved fiber coupling~\cite{meier2021comparison}. $\lambda$ can be increased either with a stronger pump laser, which is commercially available, or a more efficient source like the waveguides mentioned above. Notably, a source with these improvements is not commercial off-the-shelf technology like the source we used. Improved $D_0$ can be realized with more state-of-the art SNSPDs since some have been shown to have <1 dark count per second~\cite{shibata2015ultimate}. With more optical engineering, the mode-matching in the free-space polarization analyzers can be improved in our estimation. The most speculative of the improvements is that of $e_d$; even still, there are better polarization optics available than those employed here and more time and precision could be spent on the calibration process to improve $e_d$. Notwithstanding, it should be noted that even to achieve the current $e_d$ took significant effort and experience in the design, construction, and calibration of the polarization analyzers.
\section{Conclusion}
Here we demonstrate quantum digital signatures hardware working at low error rates between three users deployed across our campus network in three separate buildings. The current implementation 
works well enough for a proof of principle demonstration, and with realistic improvements this system could serve digital signatures to users needing low-bandwidth signature rates. To illustrate its utility, we envision a low-bandwidth but high-importance monitoring network of sensors, e.g., electrical-grid monitoring, which would benefit from additional security to protect and trust the sensor data received by the monitoring station from the deployed sensors. A quantum-digital-signatures system of sufficient bandwidth would enable increased security and trust of these sorts of deployments so the monitoring station is protected from false messages from bad actors. 

Moreover, the post-selected entanglement used here provides a very flexible low-loss way to provision entanglement to users using beamsplitters. This is only limited by the size of the beamsplitter tree used, though the raw key rate between any single pair of users will be reduced as more users are included. The current SPDC-based entangled photon source is limited by multiple pairs from creating higher rates of entanglement.  However, although they are relatively immature, replacing the present source with a polarization entangled quantum-dot-based source~\cite{Huber_2018,10.1063/5.0038729} could significantly increase the generation rate since these generally have much lower multi-pair contributions.  Alternatively, a frequency-multiplexed SPDC source could provide another more-developed path to higher rates of entanglement~\cite{Pelet_2022,Alshowkan:22,Alshowkan:22OL} though with increased complications of provisioning the entanglement compared to this work.

Note, in the QDS protocol presented in this paper, one drawback is that a relatively long raw key (on the order of $10^4$ bits) is required to sign a single bit of information. Recently, more efficient solutions for multi-bit QDS schemes have been proposed based on one-time universal hashing~\cite{10.1093/nsr/nwac228,Schiansky2023,PhysRevApplied.20.044011}. These schemes can directly generate the signature of a long message and can be more efficient than single-bit schemes. On the other hand, these schemes are built upon quantum secret sharing (QSS) protocols where classical error correction and privacy amplification steps (similar to that in QKD) are needed. An interesting future investigation could make a comparison between these two different approaches by modifying the present experimental setup into an entanglement-based QSS setup~\cite{PhysRevA.99.062311}.

\appendix
\renewcommand{\thesection}{Appendix \Alph{section}}

\section{Noise Model}
\label{App:NM}
Here we present the details of the noise model used in our simulations for our type-II distinguishable pair source.

When the source emits $\ket{0,0}$ (with probability $P(0)$),
\begin{equation}
    Y_0^Z=Y_0^X={(1-D_0)^2[1-(1-D_0)^2]}^2\approx4D_0^2\text{ and }
\end{equation}
\begin{equation}
    e_0^Z=e_0^X=0.5.
\end{equation}
Here, we ignore the terms with higher order of $D_0$.

When the source emits one photon pair (with probability $P(1)$) , the quantum state is described by $\ket{1,1}_{HV}=\frac{1}{\sqrt{2}}(\ket{2,0}_{DA}+\ket{0,2}_{DA})$. This implies an asymmetry between the H/V and D/A bases: in the H/V basis, an error occurs when the two users report detection events with the same polarization. In contrast, in the D/A basis, an error occurs when the two users report detection events with different polarization states.

After accumulating the probabilities of different paths and errors we find,
\begin{equation}
  Y_1\approx\frac{T_AT_j}{2}+(2T_A+2T_j-\frac{3}{2}T_A^2-\frac{3}{2}T_j^2)D_0+4D_0^2,
\end{equation}
\begin{equation}
    e_1^Z\approx\frac{T_A T_j e_{dz}(1-e_{dz})+(T_A+T_j-\frac{3}{4}T_A^2-\frac{3}{4}T_j^2)D_0+2D_0^2}{Y_1},\text{ and}
\end{equation}
\begin{equation}
    e_1^X\approx\frac{T_A T_j e_{dx}(1-e_{dx})+(T_A+T_j-\frac{3}{4}T_A^2-\frac{3}{4}T_j^2)D_0+2D_0^2}{Y_1}.
\end{equation}
In the above equations, we have assumed that both $T_A$ and $T_j$ are much less than one, so that all the higher order terms, such as $T_A D_0^2$, may be ignored.

When the source emits two pairs (with probability $P(2)$), the wave function can be expanded to
\begin{align}
    \ket{1,1}_{HV,1}\otimes\ket{1,1}_{HV,2}=&\text{ }\frac{1}{\sqrt{2}}(\ket{2,0}_{DA,1}+\ket{0,2}_{DA,1})\otimes\frac{1}{\sqrt{2}}(\ket{2,0}_{DA,2}+\ket{0,2}_{DA,2})\\
    =&\text{ }\frac{1}{2}(\ket{2,0}_{DA,1}\otimes\ket{2,0}_{DA,2}+\ket{2,0}_{DA,1}\otimes\ket{0,2}_{DA,2}\notag\\
    &+\ket{0,2}_{DA,1}\otimes\ket{2,0}_{DA,2}+\ket{0,2}_{DA,1}\otimes\ket{0,2}_{DA,2}).
\end{align}
For both bases, the yield is
\begin{align}
    Y_2= & \text{ }4(1-T_A-T_j)^4D_0^2+\Big(4(T_A+T_j)(1-T_A-T_j)^3+3(T_A^2+T_j^2)(1-T_A-T_j)^2+\notag\\
    &(T_A^3+T_j^3)(1-T_A-T_j)+\frac{T_A^4+T_j^4}{8}\Big)D_0+\frac{3}{2}T_AT_j(1-T_A-T_j)(2-T_A-T_j)\notag\\
    &+\frac{T_AT_j}{8}(2T_A^2+2T_j^2+3T_AT_j)
\end{align}
Now we will first treat the H/V (Z) basis then the terms in the D/A (X) basis. For the H/V basis, after accumulating the probabilities of different paths, we find the 2nd-order QBER coefficient to be
\begin{align}
     e_2^Z= & \text{ }\frac{1}{Y_2^Z}\Big(2(1-T_A-T_j)^4D_0^2+0.5[4(T_A+T_j)(1-T_A-T_j)^3\notag\\
     &+3(T_A^2+T_j^2)(1-T_A-T_j)^2+(T_A^3+T_j^3)(1-T_A-T_j)+\frac{T_A^4+T_j^4}{8}]D_0\notag\\
     &+\frac{T_AT_j}{8}[T_AT_j+4(1-T_A-T_j)(2-T_A-T_j)][1+2e_d(1-e_d)]\notag\\
     &+\frac{T_AT_j^3+T_jT_A^3}{8}[1-{e_d(1-e_d)(1-2e_d)}^2]\Big).
\end{align}
For the D/A basis, the terms $\ket{2,0}_{DA,1}\otimes\ket{2,0}_{DA,2}$ and $\ket{0,2}_{DA,1}\otimes\ket{0,2}_{DA,2}$ (each with 1/4 probability) have the same yield and QBER by symmetry; similarly, the $\ket{2,0}_{DA,1}\otimes\ket{0,2}_{DA,2}$ and $\ket{0,2}_{DA,1}\otimes\ket{2,0}_{DA,2}$ terms also have the same yield and QBER. For the former two terms, the QBER coefficient is
\begin{align}
    &e_{40}=e_{04}=\frac{1}{Y_2}\Bigg(2(1-T_A-T_j)^4D_0^2+0.5\Big(4(T_A+T_j)(1-T_A-T_j)^3\notag\\
    &+3(T_A^2+T_j^2)(1-T_A-T_j)^2+(T_A^3+T_j^3)(1-T_A-T_j)+\frac{T_A^4+T_j^4}{8}\Big)D_0\notag\\
    &+\frac{3T_AT_je_{dx}(1-e_{dx})}{4}\Big(T_AT_j+4(1-T_A-T_j)(2-T_A-T_j)\Big)\notag\\
    &+\frac{T_AT_j^3+T_jT_A^3}{8}\Big(1-(1-2e_{dx})\Big[(1-e_{dx})^3-e_{dx}^3\Big]\Big) \Bigg);
\end{align}
whereas, for the latter two terms the QBER coefficient is $e_{22}=1-e_2^Z$. This relation is due to the fact that a correct detection pattern in one basis corresponds to an error in the other basis.  Thus, overall for the D/A basis when two pairs are created the QBER coefficient is $e_2^X=(e_{40}+e_{22})/2$.

Now for $\ket{m,m}$ when $m>2$, the yield is $Y_m^X=Y_m^Z\approx2m(T_A+T_j)D_0+m^2T_AT_j$ assuming Alice's received photon number is approximately independent of Bob's (or Charlie's). Likewise, we pessimistically approximate these difficult terms $e_m^X\approx e_m^Z\approx 0.5$, assuming the polarization of Alice's received photon(s) is(are) approximately uncorrelated to that of Bob's (or Charlie's).

\section{QBER optimization algorithm}
\label{app:optdetails}
To finish calibrating the H/V QBERs and then calibrate the D/A QBERs between all three receivers across our network, we develop and employ a multi-state cyclic coordinate descent~\cite{Vassiliadis2009} algorithm, which leverages QBER optimization by curve-fitting of algorithmically determined measured samples and numerical optimization of the resulting fitted curve. As mentioned earlier, the CAS transmits the QBER after each measurement cycle. Meanwhile, the control machine has a background task that waits and collects the updated QBER, ensuring seamless integration and data synchronization. The LCVRs are connected to voltage sources of 2-kHz square waves with a variable amplitude and no DC offset (Thorlabs LCC25, Rigol DG2102, Agilent 33622A) which are connected via USB (Alice's LCVR controller is co-located with the control computer) or ethernet (Bob's and Charlie's LCVR controllers which are not co-located with the control computer).

To find the minimum QBER for a given coordinate, we employ curve fitting to minimize the impact of noise in the QBER data. To sweep through the minima for the curve fitting, two QBER samples are taken by changing the LCVR voltage amplitude by equal steps in the same direction and the slope is calculated. If the slope is negative (QBER decreasing) the algorithm continues collecting QBER samples at equally spaced voltages in that direction (voltage decreasing or increasing). If the slope is positive (QBER increasing), the slope of the voltage change is flipped so the QBER slope is decreasing. After a negative QBER slope is found, the algorithm continues collecting QBER samples at equally spaced voltages in that direction until the slope is positive again for two successive samples (and if the total samples are greater than 5, due to noise in the QBER samples) indicating we have swept through the minimum. These samples are then fit to a parabola using least-squares optimization. The fitted parabola is automatically analyzed (using numerical minimization) to find the LC voltage for the minimum QBER according to the fitted parabola. Then the LC is set to that voltage amplitude. The LC voltage-amplitude step size should be chosen to be big enough to reliably see increases or decreases given the statistical noise in each QBER sample but not too big to completely miss the QBER minimum. Due to the non-linear voltage-phase relationship of the LCVR it is necessary to change the step size for different voltage ranges.  We found step sizes of 0.1 V (0.5 V) to be good for voltages < 2V (>2 V) but a more sophisticated choice could be made using a LCVR voltage-phase calibration to ensure there is always an equal phase step.

The high-level flow of the algorithm starts by picking a QBER to minimize, then running hybrid cyclic coordinate descent on the LCVRs that affect that QBER, e.g., to start with Alice-Bob H/V QBER, minimize using Alice's and Bob's LCVRs before the NPBS. After Alice-Bob H/V QBER is minimized then attempt another QBER minimization, e.g., Alice-Charlie H/V QBER, in which case just use Charlie's LCVRs before the NPBS since Alice's were already calibrated. If there is good convergence, Bob-Charlie H/V QBER should be minimized too. If it is not, then more rounds of hybrid cyclic coordinate descent are required due to noise in the previously determined minimum voltages. Convergence will be aided by longer count times, a different step size(s), or different combinations of the LCVRs that affect a given QBER, and if desired, manual fine tuning of individual LCVRs can be done (as we did after several rounds of hybrid cyclic coordinate descent for the sake of time due to our 30-s integration time). After minimizing the H/V QBERs, the D/A QBERs can be minimized using the LCVRs after the NPBS of Alice's and Bob's receivers. The D/A basis error minimization is substantially easier at this point because it is a one-dimensional search using a single LCVR. For example, we calibrated Alice-Charlie D/A QBER using Alice's LCVR after the NPBS using the curve-fitting optimization described above. Then Alice-Bob and Bob-Charlie D/A QBERs can be averaged to be minimized using Bob's LCVR after the NPBS. At the end of this optimization, all H/V and D/A QBERs between all three pairs of users should be minimized.

\section{Simulation parameter extraction}
\label{App:paramextract}

To prepare several simulation parameters represented in Table~\ref{tab:simparams}, e.g., $\eta_0$ and $\lambda$ require additional analysis of measurement results since they were not easily directly measurable. The overall method relies on a simple model of the SPDC source combined from Ref.~\cite{TAKESUE2010276,PhysRevA.76.012307}, assuming a distinguishable entangled photon source

\begin{align}
S_A & = R \tau \sum_{n=0}^{\infty} P(n) (1-(1-D_0)(1-T_A)^n) = R\tau (1+e^{-T_A 2 \lambda} (D_0-1))\\
S_B & = R \tau \sum_{n=0}^{\infty} P(n) (1-(1-D_0)(1-T_B)^n) = R\tau (1+e^{-T_B 2 \lambda} (D_0-1))\\
C_{A,B} &= R \tau \sum_{n=0}^{\infty} P(n) (1-(1-D_0)(1-T_A)^n) (1-(1-D_0)(1-T_B)^n)\\
&= R\tau e^{-2 \lambda (T_A+T_B) }( e^{2 \lambda (T_A +T_B)}+( e^{-2 \lambda T_A}+e^{-2 \lambda T_B}) (D_0-1)+ e^{-2 \lambda T_A T_B}(D_0-1)^2),
\end{align}

Where $S_A$ and $S_B$ are the singles counts, $C_{A,B}$ is the coincidence counts, $R$ is the repetition rate, $\tau $ is the counting time, $P(n)$ is the pair generation probability distribution [Eq.~(\ref{Pnpoisson})], $D_0$ is the detector noise, and $T_i$ is the total transmission from generation to detection. 

The measured coincidences and singles counts can be compared to these calculated ones and used to estimate $\lambda$, $T_i$, and $T_j$. If there were no experimental noise, a system of three equations could be solved for the three variables. Because there is experimental noise, we use Nelder-Mead numerical optimization~\cite{2020SciPy-NMeth,Gao2012} to minimize the objective function

\begin{equation}
F = (S_A - S_A^M)^2 + (S_B - S_B^M)^2 + (C_{A,B} - C_{A,B}^M)^2,
\end{equation}

where the $M$ superscript indicates the measured values. For the initial conditions of the optimization, we use formulas from Ref.~\cite{chapphowest2018} with a first order approximation (assuming no detector noise) and solve for the pair production probability (used in Ref.~\cite{chapphowest2018} in place of $\lambda$) and the transmission of each photon path.

\begin{align}
\lambda_0 &= \frac{ S_A^M S_B^M }{R \tau C_{A,B}^M}\\
T_{A,0} &=  \frac{C_{A,B}^M}{S_B^M }\\
T_{B,0} &=  \frac{C_{A,B}^M}{S_A^M }
\end{align}

To obtain $\eta_0$, we setup the entangled photon source connected to a fiber-based polarization controller then a fiber-based polarization beam combiner and both outputs (horizontal and vertical polarization) into their own fiber-based polarization controllers then into SNSPDs. In this configuration the polarization controller was calibrated using a temporary inline polarizer to minimize the singles counts to one of the detectors to align the polarization beam combiner to the basis of the entangled photon source up to a phase. Since the source produces H/V pairs, this arrangement directly sorts the pairs of photons so each photon in the pair goes into a different detector. The efficiency of the system from the output of the entangled photon source to detection was measured to be about 0.57 per side. After measuring the singles and coincidences of the source in this arrangement, we use the method above to calculate the total transmission of each arm and back out the internal transmission of the source by dividing by 0.57. We find $\eta_0 = 0.3$.

To obtain $\lambda$, we decided it was best to use some of the measurements from the day of QDS data acquisition to avoid issues from power drifts and other differences between measurements on different days. For this, we used a representative data set; using the measured singles counts on the detectors and the coincidences for each basis, we calculated the average $\lambda$ for each user pair per basis with the methods above and averaged that together to obtain $\lambda=0.0047$ used in our simulations of the current setup.

\begin{backmatter}
\bmsection{Funding}
Content in the funding section will be generated entirely from details submitted to Prism. Authors may add placeholder text in the manuscript to assess length, but any text added to this section in the manuscript will be replaced during production and will display official funder names along with any grant numbers provided. If additional details about a funder are required, they may be added to the Acknowledgments, even if this duplicates information in the funding section.

\bmsection{Acknowledgments}
We thank Benjamin Lawrie for sharing some lab space and several single-photon detectors for the deployed fiber measurements. Also, we thank Joseph Lukens for his design contributions to the White Rabbit synchronization network and, alongside Brian Williams, to the development of the timetagger. This work was performed at Oak Ridge National Laboratory, operated by UT-Battelle for the U.S. Department of Energy under contract no. DE-AC05-00OR22725. Funding was primarily provided by the U.S. Department of Energy, Office of Cybersecurity Energy Security and Emergency Response (CESER) through the Risk Management Tools and Technologies (RMT) Program. Funding for the prior development of certain ~\ref{App:paramextract} methods was provided by U.S. Department of Energy, Office of Science, Advanced Scientific Computing Research, under the Entanglement Management and Control in Transparent Optical Quantum Networks Research programs (Field Work Proposal ERKJ378).

\bmsection{Disclosures}
The authors declare no conflicts of interest.

\bmsection{Data Availability Statement}
Data underlying the results presented in this paper are not publicly available at this time but may be obtained from the authors upon reasonable request.

\end{backmatter}



\end{document}